\documentclass[letterpaper,english]{emulateapj}
\usepackage[T1]{fontenc}
\usepackage[latin9]{inputenc}
\usepackage{amssymb}
\usepackage{graphicx}

\makeatletter

\providecommand{\tabularnewline}{\\}

\shorttitle{Characterizing the Evolving X-Ray Spectral Features During a Superburst from 4U\,1636--536}
\shortauthors{Keek, Ballantyne, Kuulkers, \& Strohmayer}

\makeatother

\usepackage{babel}
\begin{document}

\title{Characterizing the Evolving X-Ray Spectral Features During a Superburst
from 4U\,1636--536}

\author{L.~Keek and D.\,R.~Ballantyne}

\affil{Center for Relativistic Astrophysics, School of Physics, Georgia
Institute of Technology, 837 State Street, Atlanta, GA 30332-0430,
USA}

\email{l.keek@gatech.edu}

\author{E.~Kuulkers}

\affil{European Space Astronomy Centre (ESA/ESAC), Science Operations Department,
28691 Villanueva de la Cañada, Madrid, Spain}

\author{\and T.\,E.~Strohmayer}

\affil{X-ray Astrophysics Lab, Astrophysics Science Division, NASA's Goddard
Space Flight Center, Greenbelt, MD 20771, USA}
\begin{abstract}
Recent studies have shown that runaway thermonuclear burning of material
accreted onto neutron stars, i.e. Type I X-ray bursts, may affect
the accretion disk. We investigate this by performing a detailed time-resolved
spectral analysis of the superburst from 4U\,1636--536 observed in
2001 with the Rossi X-ray Timing Explorer. Superbursts are attributed
to the thermonuclear burning of carbon, and are approximately $1000$
times more energetic than the regular short Type I bursts. This allows
us to study detailed spectra for over $11\,\mathrm{ks}$, compared
to at most $100\,\mathrm{s}$ for regular bursts. A feature is present
in the superburst spectra around $6.4\,\mathrm{keV}$ that is well
fit with an emission line and an absorption edge, suggestive of reflection
of the superburst off the accretion disk. The line and edge parameters
evolve over time: the edge energy decreases from $9.4\,\mathrm{keV}$
at the peak to $8.1\,\mathrm{keV}$ in the tail, and both features
become weaker in the tail. This is only the second superburst for
which this has been detected, and shows that this behavior is present
even without strong radius expansion. Furthermore, we find the persistent
flux to almost double during the superburst, and return to the pre-superburst
level in the tail. The combination of reflection features and increased
persistent emission indicates that the superburst had a strong impact
on the inner accretion disk, and it emphasizes that X-ray bursts provide
a unique probe of accretion physics.
\end{abstract}


\section{Introduction}

Superbursts are energetic X-ray flares observed from accreting neutron
stars in low-mass X-ray binaries \citep{Cornelisse2000,Strohmayer2002}.
They are attributed to the runaway thermonuclear burning of carbon
in the ashes layer of many short Type I X-ray bursts \citep[e.g.,][]{Cumming2001,Cooper2009}.
Only 23 superbursts have been detected thus far \citep[e.g.,][]{Keek2012,Negoro2012},
some of which are only candidates, as the start of the burst is often
not observed, which makes it difficult to constrain the flare's parameters
and confirm the thermonuclear origin. Most superburst observations
have been performed with wide-field instruments with limited spectral
resolution and collecting area. The highest quality data available
is for two superburst observations with the Proportional Counter Array
(PCA) on board the \emph{Rossi X-ray Timing Explorer} (\emph{RXTE}).
One of these superbursts, from 4U~1820--30 \citep{Strohmayer2002},
was exceptionally powerful. It exhibits superexpansion during two
phases of the superburst, and it has the longest period of moderate
photospheric radius expansion (PRE) of any known X-ray burst \citep{Zand2010,Keek2012precursors}.
Detailed spectral analysis found the presence of reflection features
in the spectrum \citep{Ballantyne2004}, where part of the emission
from the neutron star surface is thought to be reflected from the
accretion disk. This allowed for the study of changes of the accretion
disk properties during an X-ray burst, which may be interpreted as
the disk receding and subsequently returning to its original configuration.
Reflection features have also been detected from neutron stars outside
of bursts \citep[e.g.,][]{Bhattacharyya2007,Cackett2008,Miller2013}.
Normal X-ray bursts last too short a time to collect enough photons
to detect reflection features with current instrumentation \citep[for constraints on the effect on the continuum for one of the brightest bursts, see][]{Zand2013}.

4U\,1636--536 is a prolific burster, that exhibits a wide range of
bursting behavior including both bursts with and without PRE \citep[e.g.,][]{Hoffman1977,Galloway2008catalog},
burst oscillations \citep{Strohmayer1998}, short recurrence time
bursts \citep{Galloway2008catalog,Keek2010}, mHz QPOs \citep{Revnivtsev2001,Altamirano2008},
double- and triple-peaked bursts \citep{Paradijs1986,Bhattacharyya2006,Zhang2009},
and four superbursts \citep{Wijnands2001sb,Strohmayer2002a,Kuulkers2009ATel}.
One superburst was observed with the PCA. Pulsations at the neutron
star spin frequency were detected near the peak of the superburst
\citep{Strohmayer2002a}. Preliminary results of the spectral analysis
show that its peak flux and its fluence are within the range that
is typical for most known superbursts \citep{2004Kuulkers,Kuulkers2003a},
such that it may be classified as a `typical' superburst, whereas
4U\,1820--30's superburst is exceptional. The start of the superburst
was observed, including a brief double-peaked precursor, although
the data quality is insufficient to establish the presence of PRE
\citep{Keek2012precursors}.

The classical approach to the spectral analysis of X-ray bursts and
superbursts is to assume that the spectrum produced by the accretion
process persists throughout the burst, i.e. the `persistent' spectrum
is unchanged. This spectrum is determined from a time interval prior
to the burst, which serves as the background during the burst. The
remainder of the spectrum is assumed to originate exclusively from
the neutron star surface, and is typically well-fit by a black body
with interstellar absorption \citep{swank1977,2002Kuulkers}, although
small deviations may be apparent in high quality spectra \citep[e.g.,][]{Paradijs1986spectra}.
A recent analysis of a large number of PRE bursts observed with the
PCA, however, finds indications that the persistent spectrum is not
constant, but has an increased normalization during the burst \citep{Worpel2013,Zand2013}.
One explanation is Poynting-Robertson drag \citep[e.g.,][]{Walker1992}:
the photons emitted by the neutron star surface transfer momentum
to the disk, reducing its angular momentum, allowing for more material
to fall in, which increases the flux generated by the accretion process.
Furthermore, stacking large numbers of X-ray bursts reveals that the
high energy ($\gtrsim30\,\mathrm{keV}$) flux is significantly reduced
during bursts, which is suggested to indicate the cooling of a corona
\citep{Maccarone2003,Chen2012,Chen2013,Ji2014}.

In this paper we present the detailed spectral analysis of the superburst
observed with the PCA from 4U~1636--536. This superburst's behavior
is a better representation of the typical superburst than 4U~1820--30,
and we investigate whether reflection features are present even for
less powerful bursts. Furthermore, we investigate changes in the persistent
flux during the superburst.

\section{Observations}

On 2/22/2001 a superburst was observed from 4U\,1636-536 with all
instruments on \emph{RXTE} \citep{Bradt1993}.\emph{ }This space-based
X-ray observatory was launched in December 1995, and operated until
January 2012. Its instrumentation consists of the PCA \citep{Jahoda2006},
the High Energy X-ray Timing Experiment \citep[HEXTE; ][]{Rothschild1998},
and the All-Sky Monitor \citep[ASM; ][]{Levine1996}. The PCA consists
of five proportional counter units (PCUs), that are sensitive to X-ray
photons in the 2 to 60~keV range, and have a combined collecting
area of $6500\,\mathrm{cm^{2}}$. During an observation only a subset
of these five may be active. The combination of its energy band and
large collecting area make the PCA the optimal instrument to study
the superburst. The ASM, although its energy band overlaps, has a
substantially smaller effective area. ASM data is, therefore, not
included in this study.

HEXTE consists of two clusters of Phoswich detectors with a band-pass
of 15 to 250~keV and an effective area of $800\,\mathrm{cm^{2}}$
per cluster. During an observation the clusters alternately point
on and off source, such that the source is always observed by one
cluster, while the other measures the high-energy background. The
HEXTE light curve of this superburst has been previously presented
by \citet{Kuulkers2010}. We will briefly mention how the HEXTE spectral
data compares to PCA data, but our analysis will focus on the latter.
We extract HEXTE spectra from Standard 2 data for each cluster, separating
the on-source and off-source pointings. We check that no bright X-ray
source was present in the field of view at the off-source positions.
Using the tool \texttt{hxtdead}, we correct for `dead time': the reduction
of the exposure by the time that the on-board electronics takes to
process a detected event. During the analysis, the off-source spectrum
is subtracted as a background from the on-source spectrum.

Due to an overflow of the on-board data buffer some data products
were lost, most notably the high time resolution PCA data at the superburst
onset \citep{Strohmayer2002a}. Throughout the entire superburst Standard
1 and 2 data products are available. The former has a time resolution
of $\frac{1}{8}\,\mathrm{s}$, but no spectral information, whereas
the latter has a time resolution of $16\,\mathrm{s}$ and $129$ energy
channels. From the Standard 1 light curve we find that, apart from
the very start of the superburst, the Standard 2 time resolution is
sufficient for our purposes.

The superburst observation lasted $4$ subsequent \emph{RXTE} orbits
(Obs IDs 50030-02-08-01 and 50030-02-08-02), and is interrupted by
$3$ data gaps due to Earth occultations. For time-resolved spectroscopy
during the first $2$ orbits we choose a time resolution of $64\,\mathrm{s}$,
which increases the statistics of the spectra over the maximal resolution
of $16\,\mathrm{s}$, while still providing sufficient sampling of
the changes in the spectral parameters with time \citep[see also][]{Strohmayer2002}.
In the last $2$ orbits the count rate per PCU is lower by over a
factor $4$ compared to the peak. We increase the time resolution
 to $128\,\mathrm{s}$ to obtain spectra with similar statistics.
This resolution is still sufficient to follow the changes with time,
because in this part of the superburst tail the decay of the light
curve has slowed down.

We extract spectra from data in the $3$ to $20\,\mathrm{keV}$ energy
range, which is well calibrated, whereas at higher energies the instrument
background dominates, and uncertainties in the background modeling
may be substantial \citep{Jahoda2006}. Different combinations of
PCUs were active during the different orbits. We use PCUs 0, 2, and
3, which were on during all four orbits. Using the procedure recommended
by the instrument team,%
\footnote{\texttt{\scriptsize{http://heasarc.gsfc.nasa.gov/docs/xte/recipes/pcadeadtime.html}}%
} we correct for dead time. Furthermore, we group neighboring channels
if the number of counts of a channel is less than $15$, ensuring
that $\chi^{2}$ statistics are applicable.

We use the tool \texttt{pcabackest} to model the instrument background,
which is based on blank-sky observations and takes into account a
possible particle background from prior passage through the South
Atlantic Anomaly. We generate one background spectrum per orbit. The
background is, however, expected to change somewhat during an orbit
\citep{Jahoda2006}. In tests where we generate a background for each
$64\,\mathrm{s}$ or $128\,\mathrm{s}$ time interval, the results
of our analysis did not change significantly, because the orbital
variability is strongest $\gtrsim20\,\mathrm{keV}$, which is outside
the considered energy range. We give preference to the single background
spectrum per orbit, because the background model uses the anti-coincident
signals, and its relative error is reduced when a longer time interval
is used.

Spectral analysis is performed with \texttt{XSPEC} version 12.8.1
\citep{Arnaud1996}. A $0.5\%$ systematic uncertainty is added in
quadrature to the errors in the data points of the spectra, to take
into account uncertainties in the response \citep{Jahoda2006}. The
reported uncertainties are at $1\sigma$.

At the start of the observation, the spacecraft was slewing. We only
consider data from the time when the final pointing was reached. This
is of no consequence to our analysis, because the source appeared
to have a constant photon flux during and directly after the slew.
We define the time $t=0$ at MJD~$51962.702069$.

Finally, near the peak of the superburst, oscillations have been detected
with an amplitude of $1\%$ at the neutron star's spin frequency of
$581.9\,\mathrm{Hz}$ \citep{Strohmayer2002a}. This suggests an inhomogeneity
in the heating of the star's surface. Because we integrate spectra
over much longer time intervals than the spin period, and because
of the small amplitude, we do not expect the oscillations to significantly
alter the spectral shape.

\section{Results}

First we follow the classical approach to X-ray burst and superburst
spectroscopy by subtracting a persistent spectrum and fitting the
net spectrum with an absorbed black body. Next we attempt to improve
the spectral fits by investigating a changing persistent spectrum
as well as emission and absorption features.

\subsection{Persistent emission: fit to pre-superburst orbit\label{sub:Persistent-emission:-fit}}

We investigate the persistent flux in the orbit immediately prior
to the superburst observation. In this orbit the count rate is comparable
to that at the end of the superburst tail. We extract a spectrum
using data from the entire orbit. First we fit a cut-off power law
model that depends on energy, $E$, as $K_{\mathrm{pl}}E^{-\Gamma}\mathrm{e}^{-E/E_{\mathrm{cutoff}}}$
with normalization $K_{\mathrm{pl}}$, photon index $\Gamma$, and
a high-energy exponential cutoff at $E_{\mathrm{cutoff}}$ (model
\texttt{cutoffpl} in XSPEC). We include photoelectric absorption using
the \texttt{vphabs} model and fix its parameters, including the hydrogen
column $N_{\mathrm{H}}$, to the mean of the best fit values of \citet{Pandel2008}
(Table~\ref{tab:Fpers}; abundances relative to \citealt{Wilms2000};
cross sections from \citealt{Balucinska1992}). The best fit has for
the goodness of fit per degree of freedom $\chi_{\nu}^{2}=3.4$. 

Including a black body component at lower energies does not substantially
improve the fit, nor does allowing $N_{\mathrm{H}}$ to vary. Including
a (smeared) absorption edge gives $\chi_{\nu}^{2}=1.5$ with an apparent
feature in the fit residuals around $6.4\,\mathrm{keV}$. Including
instead a Gaussian emission line with centroid energy $E_{\mathrm{line}}$,
width $\sigma_{\mathrm{line}}$, and normalization $K_{\mathrm{line}}$
yields $\chi_{\nu}^{2}=0.75$ without strong features in the residuals.
When $N_{\mathrm{H}}$ is allowed to vary in the latter model with
a Gaussian, we find $\chi_{\nu}^{2}=0.73$ and a vanishingly small
value for $N_{\mathrm{H}}$ that is consistent with $0$ within its
$1\sigma$ uncertainty. Therefore, we choose \texttt{vphabs(cutoffpl+gaussian)}
with $N_{\mathrm{H}}$ fixed to the mean value from \citet{Pandel2008}
as our fiducial model for the persistent spectrum (Table~\ref{tab:Fpers}).
This model is consistent with the best fit model found in other, broad-band
studies \citep[see Section \ref{sub:The-persistent-spectrum}]{Pandel2008,Lyu2014}.
With this model we determine the unabsorbed flux of the cut-off power
law component both in the $3$ to $20\,\mathrm{keV}$ energy band,
$F_{3-20}$, and extrapolated to the $0.03$ to $100\,\mathrm{keV}$
band, $F_{0.03-100}$, as a measure of the bolometric flux (Table~\ref{tab:Fpers}).
The Gaussian line has an equivalent width of $0.10\pm0.02\,\mathrm{keV}$.

Simultaneous fits of the PCA and HEXTE data give results consistent
with PCA-only fits. The HEXTE spectra confirm the cut-off power law
continues in the energy range $15-35\,\mathrm{keV}$. At higher energies
the number of counts is rather low and the background dominates the
spectrum; no significant deviation from the cut-off power law is observed
at these energies.

Using the same model we also analyze the spectrum of the orbit that
starts $8$~hours after the superburst onset. The flux is close to
$4\%$ higher than the pre-superburst flux. 

\begin{table}
\caption{\label{tab:Fpers}Best fit to pre- and post-superburst persistent
spectrum$^{\mathrm{a}}$ }

\begin{tabular}{cll}
\hline 
 & pre-superburst & post-superburst\tabularnewline
Obs Id & 50030-02-08-00 & 50030-02-08-03\tabularnewline
\hline 
\texttt{vphabs } & \multicolumn{2}{c}{fixed \citep{Pandel2008}}\tabularnewline
\hline 
$N_{\mathrm{H}}\,(10^{22}\mathrm{cm^{-2}})$ & \multicolumn{2}{c}{$0.379$}\tabularnewline
O & \multicolumn{2}{c}{$1.29$}\tabularnewline
Ne & \multicolumn{2}{c}{$1.4$}\tabularnewline
Si & \multicolumn{2}{c}{$1.5$}\tabularnewline
Fe & \multicolumn{2}{c}{$1.48$}\tabularnewline
\hline 
\texttt{cutoffpl} &  & \tabularnewline
\hline 
$\Gamma$ & $1.06\pm0.02$ & $1.22\pm0.02$\tabularnewline
$E_{\mathrm{cutoff}}\,(\mathrm{keV})$ & $4.80\pm0.06$ & $5.03\pm0.07$\tabularnewline
$K_{\mathrm{pl}}\,(\mathrm{c\, s^{-1}\, cm^{-2}\, keV^{-1}})$ & $1.24\pm0.02$ & $1.65\pm0.03$\tabularnewline
$F_{3-20}\,(\mathrm{10^{-9}erg\, s^{-1}\, cm^{-2}})$  & $4.520\pm0.006$ & $4.693\pm0.007$\tabularnewline
$F_{0.03-100}\,(\mathrm{10^{-9}erg\, s^{-1}\, cm^{-2}})$  & $9.037\pm0.010$ & $10.8\pm0.2$\tabularnewline
\hline 
\texttt{gaussian} &  & \tabularnewline
\hline 
$E_{\mathrm{line}}\,\mathrm{(keV)}$ & $6.67\pm0.13$ & $6.8\pm0.2$\tabularnewline
$\sigma_{\mathrm{line}}\,(\mathrm{keV})$ & $0.87\pm0.14$ & $0.8\pm0.2$\tabularnewline
 $K_{\mathrm{line}}\,(10^{-3}\mathrm{c\, s^{-1}\, cm^{-2}\, keV^{-1}})$ & $4.3\pm0.8$ & $2.9\pm0.9$\tabularnewline
Equivalent width $(\mathrm{keV})$ & $0.10\pm0.02$ & $0.07\pm0.02$\tabularnewline
\hline 
$\chi_{\nu}^{2}\,(\mathrm{degrees\, of\, freedom})$  & $0.75\,(33)$ & $0.79\,(33)$\tabularnewline
\hline 
\end{tabular}

$^{\mathrm{a}}$ See Section~\ref{sub:Persistent-emission:-fit}
for a description of the model components and their parameters
\end{table}

\subsection{Classic superburst spectral fits\label{sub:Classical-superburst-spectral}}

\begin{figure}
\includegraphics{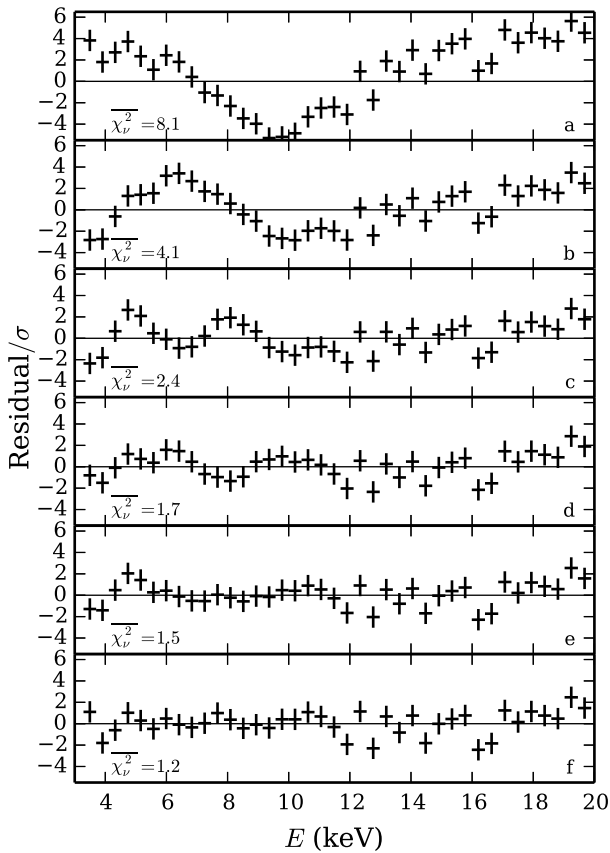}

\caption{\label{fig:Fit-residuals}From spectral fits to a $64\,\mathrm{s}$
spectrum at $t=1600\,\mathrm{s}$, the fit residuals in units of the
uncertainty of each data point, $\sigma$, for several models: \textbf{a}:
absorbed (fixed $N_{\mathrm{H}}$) black body with pre-superburst
persistent spectrum subtracted; \textbf{b}: model \emph{a} with variable
normalization of persistent flux; \textbf{c}: \emph{b} + Gaussian
emission line; \textbf{d}: \emph{b} + absorption edge; \textbf{e}:
\emph{b} + line and edge; \textbf{f}: \emph{e} + variable $N_{\mathrm{H}}$.
For each model we indicate $\overline{\chi_{\nu}^{2}}$: the mean
$\chi_{\nu}^{2}$ in the first orbit.}
\end{figure}

Extracting spectra every $64\,\mathrm{s}$ in the first two and $128\,\mathrm{s}$
in the last two orbits, we follow the common approach of deriving
the net superburst spectra by subtracting the pre-superburst persistent
spectrum. To this we fit a black body with temperature $kT$ and normalization
$K_{\mathrm{bb}}\equiv R^{2}d^{-2}$, with radius of the emitting
area $R$ in km and distance $d$ in units of $10$~kpc (XSPEC model
\texttt{bbodyrad}), and we take into account interstellar absorption
as described in the previous section. The best fit has an average
$\chi_{\nu}^{2}$ of $\overline{\chi_{\nu}^{2}}=8.1$ in the first
orbit where the data quality is highest. There is a strong feature
in the fit residuals around $7\,\mathrm{keV}$, and the high-energy
tail has a higher flux compared to the best fit black body (Figure~\ref{fig:Fit-residuals}a).

\subsection{Variable persistent flux and features in the first orbit\label{sub:orbit 1}}

\begin{figure*}
\begin{centering}
\includegraphics{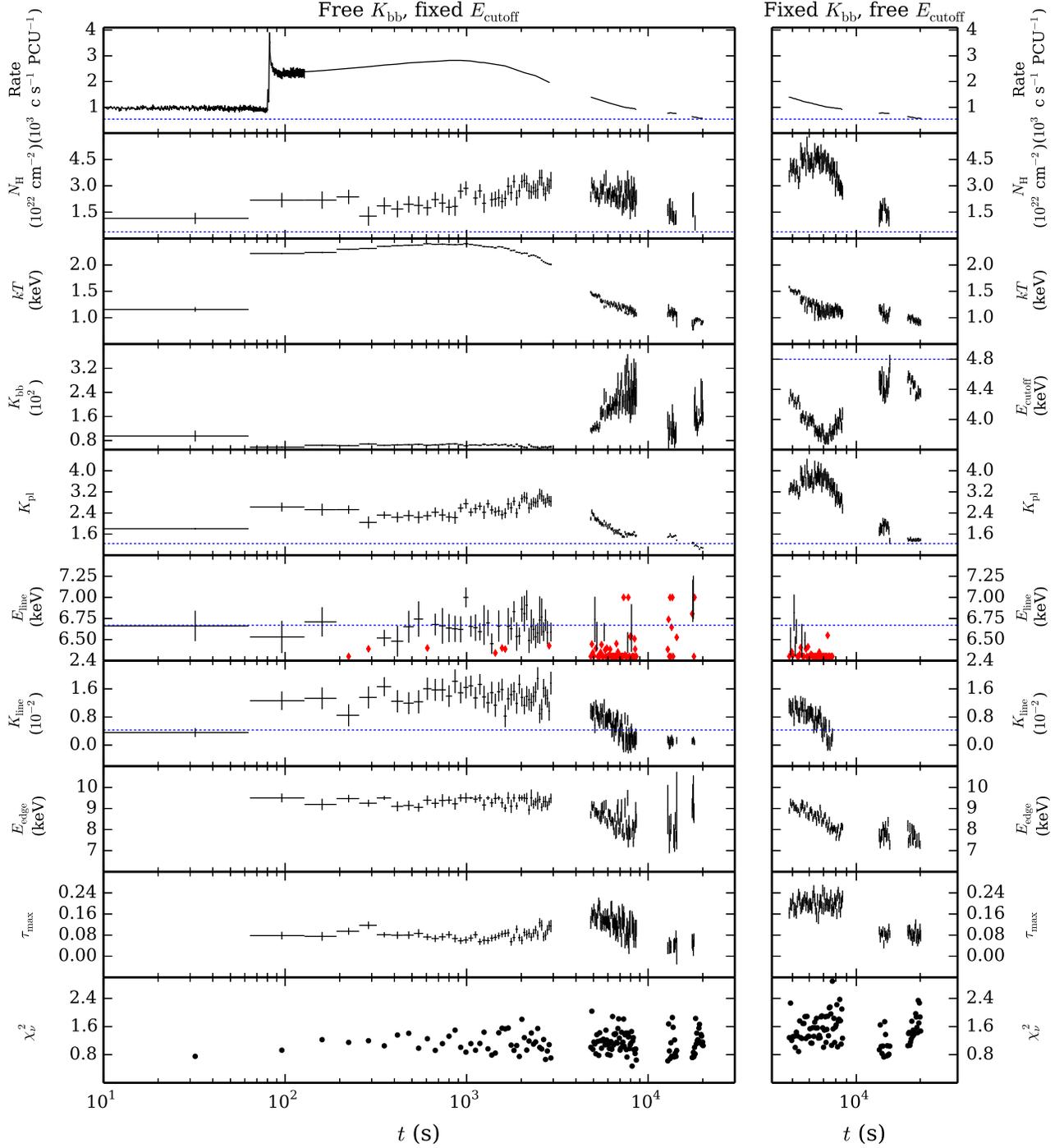}
\par\end{centering}

\caption{\label{fig:best fit}Model parameters of spectral fits as a function
of time, $t$. \textbf{Left:} top panel shows the count rate from
the Standard 1 data. The first $128\,\mathrm{s}$ has the maximum
time resolution of $\frac{1}{8}\,\mathrm{s}$; for later times the
time bins are as used in the spectral analysis. Other panels show
the best fit values of the parameters for the model that best fits
the first orbit (Section~\ref{sub:orbit 1}). Each data point indicates
its $1\sigma$ uncertainty vertically as well as the width of the
time interval horizontally. Where $E_{\mathrm{line}}$ is ill-constrained,
we only indicate the value by a diamond. Dotted lines indicate values
from the pre-superburst orbit (Table~\ref{tab:Fpers}). The normalizations
of the model components, $K_{\mathrm{X}}$, have units $\mathrm{c\, s^{-1}\, cm^{-2}\, keV^{-1}}$,
with an optional prefactor indicated in the label. \textbf{Right}:
best fit to the last $3$ orbits with $K_{\mathrm{bb}}$ fixed to
the mean value from the first orbit (left panel) and with free $E_{\mathrm{cutoff}}$
(Section~\ref{sub:orbit 2}).}
\end{figure*}

To investigate improvements to the spectral fits, we start by considering
the first orbit, where the count rate and data quality are highest.
We no longer subtract the persistent spectrum. Instead, we subtract
a modeled instrumental background, and include in the fits the persistent
spectral model (Table~\ref{tab:Fpers}) in addition to a black body.
The parameters of the persistent model, a cut-off power law, are fixed
to the pre-superburst best-fit values, with the exception of $K_{\mathrm{pl}}$
\citep[similar to the approach by][]{Worpel2013}. Also, we leave
out the Gaussian emission line from the persistent model, as it is
undetectable in the short time intervals during the burst. The best
fit improves: $\overline{\chi_{\nu}^{2}}=4.1$, and the excess at
high energies is reduced. The feature around $7\,\mathrm{keV}$, however,
remains (Figure~\ref{fig:Fit-residuals}b). Allowing more parameters
of the cut-off power law to vary during the fit further improves $\chi_{\nu}^{2}$,
but competition between the power law and the black body components
leads to large uncertainties in the fit parameters. We, therefore,
choose to keep all cut-off power law parameters fixed except $K_{\mathrm{pl}}$.

Inspired by previous studies \citep{Strohmayer2002,Ballantyne2004},
we investigate the $\sim7\,\mathrm{keV}$ feature's interpretation
as a reflection line and/or an absorption edge. We employ as additional
model components a \texttt{gaussian} and an \texttt{edge}. Note that
we consider this Gaussian to be different from the one in the pre-superburst
spectrum, and scale it independently from the cut-off power law. When
the parameters of these components are left unconstrained, the line
and edge may `drift' away from the location of the feature and compete
with the two continuum components. We, therefore, restrict certain
parameters to a physically motivated range: the Gaussian line's energy
to $6.3\,\mathrm{keV}\leq E_{\mathrm{line}}\leq7.0\,\mathrm{keV}$
and width $\sigma=0.2\,\mathrm{keV}$, and the edge energy to $7.0\,\mathrm{keV}\leq E_{\mathrm{edge}}\leq9.5\,\mathrm{keV}$.
Note that if we allow the Gaussian's width to vary freely, it will
fit to the continuum instead of the feature. Because of the limited
spectral resolution, the precise value of $\sigma$ is of little consequence
to the best-fit values of the other parameters. Furthermore, we analyze
spectra in chronological order, and use the best fit values of the
previous interval as starting values for the fit of the next interval,
such that we can follow changes over time even when the significance
of these features is reduced in the superburst tail.

Including either the Gaussian line (Figure~\ref{fig:Fit-residuals}c)
or the edge (Figure~\ref{fig:Fit-residuals}d) improves the fit:
$\overline{\chi_{\nu}^{2}}=2.4$ and $\overline{\chi_{\nu}^{2}}=1.7$,
respectively. Features remain, however, in the residuals. Adding both
components reduces these features and further improves the fit: $\overline{\chi_{\nu}^{2}}=1.5$
(Figure~\ref{fig:Fit-residuals}e).

The most pronounced remaining issue in the residuals is at the lowest
energies $E\lesssim5\,\mathrm{keV}$. Allowing $N_{\mathrm{H}}$ of
the absorber to vary, removes this: $\overline{\chi_{\nu}^{2}}=1.2$.
There are no more significant features visible in the residuals (Figure~\ref{fig:Fit-residuals}f).
The measured distribution of the $\chi^{2}$ values of all time bins
in the first orbit is consistent with the theoretical $\chi^{2}$
distribution for perfect fits.

In conclusion, the spectra in the first orbit are well described by
the combination of an absorbed black body and cut-off power law, where
the shape of the cut-off power law is fixed to the pre-superburst
persistent values, whereas its normalization is left free. In addition,
a Gaussian, an absorption edge, and increased values of $N_{\mathrm{H}}$
describe spectral features at lower energies. We apply this model
(\texttt{vphabs{*}edge(cutoffpl+bbodyrad+gaussian)} in XSPEC) to all
spectra from the entire superburst (Figure~\ref{fig:best fit} left).

Comparing $K_{\mathrm{line}}$ from the pre-superburst spectrum to
the weighted mean from the first superburst orbit, the latter is larger
by a factor $3.0\pm0.6$. Given that the spectra during the superburst
have fewer counts than the pre-superburst spectrum, this justifies
our choice to include only one (dominant) Gaussian line in our model.

The first $80\,\mathrm{s}$ of our data shows a plateau of enhanced
flux (Figure~\ref{fig:best fit} top left), which we include in our
fits. Attempting a fit with only an absorbed cut-off power law, similar
to the pre-superburst fits, leaves an excess at low energies that
is not well-fit by a gaussian line (with $\sigma<1\,\mathrm{keV}$),
but that is well-described by a black body.

\subsection{Evolution of the spectrum in the second orbit\label{sub:orbit 2}}

Applying the best fit model for the first orbit to the second orbit,
we find an increase in the black-body normalization (Figure~\ref{fig:best fit}
left), implying a mean radius expansion by a factor $2.36\pm0.03$.
PRE usually occurs at the peak of energetic bursts, not in the tail.
Our model, therefore, most likely does not correctly describe the
physical behavior of the system (see Section~\ref{sub:Black-body-emitting-area}
for alternative explanations). 

We take the mean value of the black-body normalization from the first
orbit, $K_{\mathrm{bb}}=62.3\,\mathrm{c\, s^{-1}\, cm^{-2}\, keV^{-1}}$,
and for fits to the subsequent orbits fix it to this number. This
does not provide a good fit, which suggests that the shape of the
cut-off power law component changes from the pre-superburst value.
Indeed, allowing $E_{\mathrm{cutoff}}$ to vary produces reasonable
fits: $\chi_{\nu}^{2}=1.6$ on average in the second orbit. When we
allow $\Gamma$ to vary instead, similar improvements to the fits
are obtained. One would like to further improve the fits to optimize
$\chi^{2}$. Allowing one extra parameter from the black-body or power-law
model components to vary indeed produces $\chi^{2}$ values that follow
the ideal distribution. As mentioned previously, competition between
these two components causes large uncertainties in the fit parameters.
We, therefore, can use only a limited number of free parameters to
fit the continuum, and we cannot simultaneously constrain both the
black body normalization and the power law shape.

During the superburst decay, $E_{\mathrm{cutoff}}$ is substantially
lower than measured in the pre-superburst orbit. At $t=7.2\times10^{3}\,\mathrm{s}$
a minimum is reached of $E_{\mathrm{cutoff}}=3.74\pm0.03\,\mathrm{keV}$
(determined from the weighted mean of $4$ bins around the maximum;
Figure~\ref{fig:best fit} right). This is preceded at $t=6.5\times10^{3}\,\mathrm{s}$
by a maximum in $N_{\mathrm{H}}=(4.7\pm0.2)\times10^{22}\,\mathrm{cm^{-2}}$.

If we apply the same fit to the first orbit, $E_{\mathrm{cutoff}}$
has a somewhat larger value than pre-superburst. The uncertainty in
each data point, however, is relatively large, because around the
superburst peak the spectrum is dominated by the black body. Therefore,
the fit here is not strongly dependent on the precise value of $E_{\mathrm{cutoff}}$.
We continue to use the best fit model from Section~\ref{sub:orbit 1}
in orbit 1, and we employ the model with fixed $K_{\mathrm{bb}}$
and free $E_{\mathrm{cutoff}}$ in the subsequent orbits.

\begin{figure*}
\begin{centering}
\includegraphics{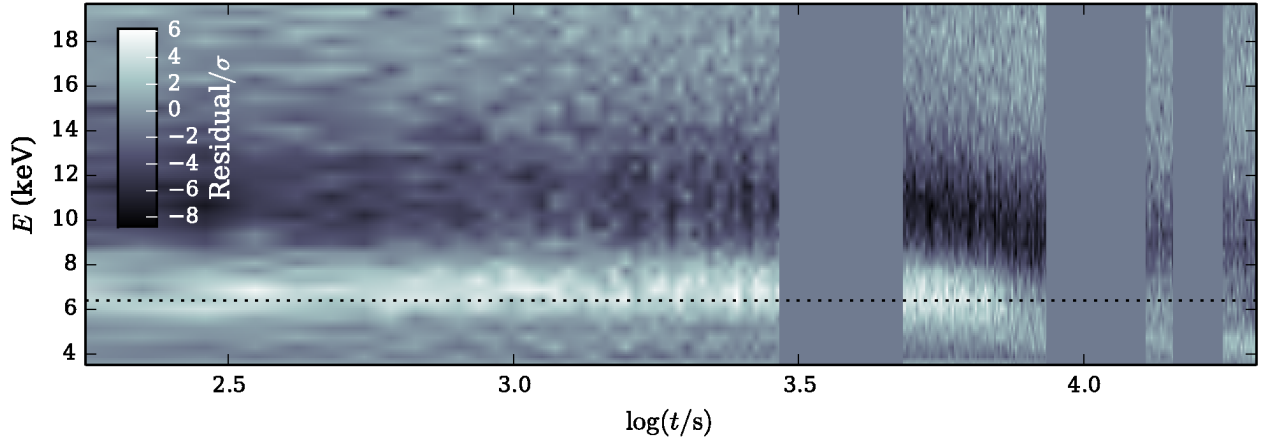}
\par\end{centering}

\caption{Fit residuals as a multiple of $\sigma$ in all energy channels between
$3\,\mathrm{keV}$ and $20\,\mathrm{keV}$ for all time intervals
during the superburst. The fitted model is our best fit model (Section~\ref{sub:orbit 1}
and \ref{sub:orbit 2}). In the first orbit we have a fixed $E_{\mathrm{cutoff}}$
and free $K_{\mathrm{bb}}$, whereas in the subsequent orbits we fix
the latter while allowing $E_{\mathrm{cutoff}}$ to vary. Data gaps
are represented by areas of uniform color. A horizontal dotted line
is placed at $6.4\,\mathrm{keV}$\label{fig:residuals-img}}
\end{figure*}
 The uncertainty in the line and edge parameters for each data point
is rather large. $E_{\mathrm{line}}$ is not well-constrained in most
of the time bins of the second orbits, although $K_{\mathrm{line}}$
is constrained, because we restricted $E_{\mathrm{line}}$ to an energy
range that lies within the broad spectral feature around $7\,\mathrm{keV}$.
To illustrate the clear presence of this feature, its shift in energy,
and its reduction in amplitude over time, we plot the fit residuals
as a function of time (Figure~\ref{fig:residuals-img}). For this,
we use the two best fit models for the first and subsequent orbits,
respectively, without the edge and line components.

\subsection{Return to pre-superburst values in the third and fourth orbits}

The line is not detected and the edge is detected at lower significance
in the final two orbits of the superburst observation, even though
we integrate our spectra over $128\,\mathrm{s}$. $N_{\mathrm{H}}$
is reduced in orbit $3$ compared to orbits $1$ and $2$. In the
last orbit it is no longer constrained by the data, similar to the
fits of the pre-superburst data (Section~\ref{sub:Persistent-emission:-fit}).
Furthermore, the cut-off power law parameters approach the persistent
values.

\subsection{Black-body and power-law flux}

\begin{figure*}
\begin{centering}
\includegraphics{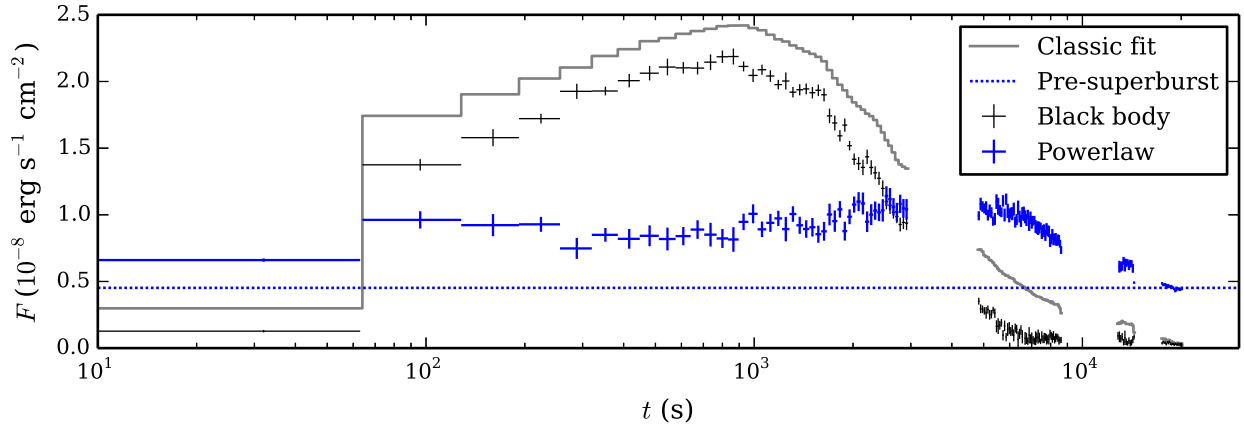}
\par\end{centering}

\caption{\label{fig:flux}Unabsorbed $3$ to $20$~keV fluxes, $F$, as a
function of time, $t$. We show both the black-body flux from the
classic fit (Section~\ref{sub:Classical-superburst-spectral}) and
the pre-superburst persistent flux (Section~\ref{sub:Persistent-emission:-fit}),
as well as the results of simultaneous fits with the black body and
cutoff power law components. For the latter we employ the model with
free $K_{\mathrm{bb}}$ and fixed $E_{\mathrm{cutoff}}$ in the first
orbit, and vice versa in the later orbits. }
\end{figure*}
 The classic fit yields a bolometric unabsorbed black body fluence
that is larger than the one from our ``optimal'' fits (Section~\ref{sub:orbit 2})
by a factor $1.72\pm0.12$ . For the data gaps we perform linear interpolation,
using the mean of three data points on each side of the gaps. Without
interpolation, the factor is $1.49\pm0.10$.

In the first orbit, the in-band flux is dominated by the black body,
whereas the power law dominates at later times (Figure~\ref{fig:flux}).
Furthermore, only in the first orbit, around the count rate peak,
is the peak of the black body distribution of counts within the observed
energy range. In the subsequent orbits the peak is at an energy that
is lower than the considered energies, and only the tail is visible.
The peak of the black body energy distribution is always well within
the considered energy range, such that we capture most of the black
body energy flux.

We obtain high values of $N_{\mathrm{H}}$ for the neutral absorption
component of our spectral model. This suggests that a substantial
part of the black-body flux is absorbed. In the $3$ to $20\,\mathrm{keV}$
band the weighted mean of the fraction of the black-body flux that
is absorbed is $5.1\%\pm0.9\%$, $13\%\pm4\%$, and $5\%\pm2\%$ in
orbits $1$, $2$, and $3$, respectively. The neutral absorber is,
however, constrained by only a few spectral bins at low energy. Detailed
spectral modeling of both an ionized absorber and the reflection features
may provide a better motivated description of this part of the spectrum.

\section{Discussion}

We have presented the results of our time resolved spectral analysis
of the \emph{RXTE }PCA spectrum of the 2001 superburst from 4U\,1636-536.
We find features in the spectra of both the superburst and pre-superburst
emission that are suggestive of reflection off the accretion disk.
Furthermore, the persistent flux more than doubles during the superburst,
and returns to the pre-superburst value in the tail.

\subsection{Components of the persistent spectrum}

\label{sub:The-persistent-spectrum} We have analyzed the spectrum
of the persistent flux in the orbit directly prior to the superburst
and in an orbit after the superburst (Table~\ref{tab:Fpers}). The
flux in the latter orbit is $4\%$ larger, which may be due to the
tail of the superburst, but variations in the persistent flux of this
magnitude are common in 4U\,1636--536 on the timescale of the \emph{RXTE
}orbit. Both spectra are well described by an absorbed cut-off power
law in combination with a Gaussian emission line. This is consistent
with previous analyses of higher resolution spectra from observations
with \emph{XMM-Newton} and \emph{Suzaku} \citep{Pandel2008,Lyu2014}.
These spectra extend to lower energies ($\sim0.5\,\mathrm{keV}$),
and discern two extra thermal components, which are attributed to
the inner accretion disk and the neutron star surface. The latter
possibly includes a contribution from a boundary layer, depending
on where the accreted material dissipates its angular momentum as
it approaches the neutron star surface. The thermal components have
temperatures $kT\lesssim1\,\mathrm{keV}$, such that their contribution
to the flux in the PCA band is negligible \citep{Pandel2008,Lyu2014}.
The cut-off power law component is thought to be produced by Compton
scattering off hot electrons in a corona. Irradiation of the inner
accretion disk by the corona and the neutron star surface may produce
a reflection spectrum, which includes an emission line \citep[e.g.,][]{Miller2007}.
Our spectral model, therefore, qualitatively agrees with the previous
studies, and is quantitatively similar, keeping in mind the time variability
of the persistent flux \citep[e.g.,][]{Pandel2008}. \citet{Lyu2014}
identify correlations between the different spectral components as
a function of the hardness of the source, but it may not be correct
to directly compare these to the behavior that we observe during the
superburst, as the processes that set the state of the disk are different.

During the superburst the neutron star surface heats up, and its thermal
emission dominates the PCA spectrum at lower energies. The cut-off
power-law flux in the PCA band approximately doubles. If the black-body
flux from the inner accretion disk and the boundary layer also double,
then the black-body temperatures increase by only $19\%$, or even
less if the emitting area grows as well. The contribution of these
two thermal components to the flux in the PCA band is, therefore,
expected to remain small compared to the main black body component
during the superburst.

\subsection{Reflection features}

Flux reflected off the accretion disk can produce a reflection spectrum
that includes a relativistically broadened fluorescent Fe K$\alpha$
emission line as well as a blend of absorption edges \citep[for a review see, e.g.,][]{Miller2007}.
Indeed, an emission line close to $6.4\,\mathrm{keV}$ is detected
from 4U\,1636--536 \citep{Pandel2008,Lyu2014}. We also find this
feature in the pre-superburst spectrum, although the edge is not detected,
because the feature is rather weak. Because outside of the burst the
cut-off power law dominates the flux, this line may be interpreted
as reflection of the flux from the corona. The line has a large width,
which may in part be explained by a blending of Fe lines from different
ionization states \citep{Pandel2008}.

During the superburst we need both an emission line and an absorption
edge to obtain satisfactory fits. A similar requirement was found
for the 1999 superburst from 4U\,1820--30 \citep{Strohmayer2002},
and these features were found to be well described by reflection of
the neutron star surface emission off the accretion disk \citep{Ballantyne2004}.
The line as observed during the superburst is much stronger than if
it were to be produced by reflection of the brightened cut-off power
law, and is therefore predominantly powered by reflected burst emission
from the neutron star surface. Note that local spectral features have
been detected with the PCA in bright X-ray bursts that were not well-fit
with a reflection model, but are better described by absorption edges
thought to originate with heavy element ashes mixed into the neutron
star photosphere \citep{Zand2010}. Therefore, despite the limited
spectral resolution of the PCA, it is possible to distinguish reflection
features from absorption features in the neutron star photosphere.

Comparing the line and edge parameters in the first \emph{RXTE} satellite
orbit of the superbursts from 4U\,1820--30 and 4U\,1636--536, in
both cases the line and edge energies decrease somewhat during the
burst, and so do the line normalization and the edge depth \citep[this paper]{Strohmayer2002}.
For 4U\,1820--30 it was argued that the receded inner disk returns
in the tail to smaller radii, increasing the gravitational red shift
\citep{Ballantyne2004}. Also, for that superburst the reflection
features are somewhat stronger, which is consistent with the reflection
of emission from a brighter superburst.

In a future study we will further investigate these spectral features
using detailed reflection spectra, similar to the study by \citet{Ballantyne2004}.

\subsection{Superburst increases persistent emission}

$K_{\mathrm{pl}}$ in the first \emph{RXTE} orbit is on average larger
than measured before the superburst by a factor $1.83\pm0.03$ \citep[the $f_a$ factor in][]{Worpel2013},
and the cut-off power-law flux increases with this factor as well.
In the superburst tail, the cut-off power-law flux smoothly returns
to the pre-superburst value. \citet{Worpel2013} infer much larger
increases for a selection of PRE bursts observed with the PCA, including
from 4U\,1636--536 \citep[see also][]{Zand2013}. This may be because
the superburst is less bright, as it does not reach the Eddington
limit.

The increase in the cut-off power-law flux has been suggested to result
from an increase in accretion due to Poynting-Robertson drag \citep[e.g.,][]{Walker1992}
induced by the burst emission from the neutron star surface \citep{Worpel2013}.
Alternatively, the enhanced cut-off power-law could be the result
of an increase in seed photons for the Comptonization process \citep{Zand2013}.
These seed photons would be produced by the reprocessing in the accretion
disk of the burst emission. The reflection features may trace the
irradiation of the disk and, therefore, the reprocessed flux. The
increase in the normalization is, however, much larger for the Gaussian
line than for the cut-off power law. This may be due to a difference
in the efficiency for reflection and reprocessing. Detailed modeling
of the reflection spectrum can provide further insight.

Recent X-ray burst atmosphere spectral models provide a better physical
description of the neutron star photosphere, and they predict a hard
comptonized tail in addition to the black body \citep{Suleimanov2010}.
This tail, however, is substantially weaker than the cut-off power
law component in our model. In fact, in the presence of variable persistent
emission, such as what we find in the superburst, fits with the atmosphere
model are problematic \citep{Suleimanov2011}. We have tested the
use of the atmosphere model as a substitute for the black-body component,
and we find that $\chi^{2}$ does not improve. The increase in the
cut-off power law is also too strong to be explained by rotational
Doppler broadening.

It is interesting that no variable persistent component was required
to describe the spectra of the 1999 superburst from 4U\,1820--30,
for which similar data products are available as for the superburst
studied in this paper \citep{Strohmayer2002,Ballantyne2004}. That
superburst was exceptionally bright and it reached the Eddington limit
\citep{Strohmayer2002,Zand2010,Keek2012precursors}. If the enhanced
cut-off power-law flux is due to Poynting-Robertson drag induced accretion,
the effect is expected to be stronger for 4U\,1820--30 than for 4U\,1636--536.
Perhaps  the Eddington limited flux from the neutron star during
superexpansion as well as the long period of moderate radius expansion
prevent an increase in the accretion rate \citep[e.g.,][]{Ballantyne2005}.
Both the occurrence of superexpansion, during which the flux drops
below the pre-superburst level, and of variability in the tail \citep[achromatic in the PCA-band; see][]{Kuulkers2003a,Zand2011}
may be indications of mass outflow during the 4U\,1820--30 superburst.
The 4U\,1636--536 superburst lacks both features.

\subsection{Black-body emitting area and the power law cut-off}

\label{sub:Black-body-emitting-area} The PCA spectra constrain only
a limited number of parameters, and we must make assumptions about
the rest. The assumption of a fixed shape of the cut-off power law
produces in the second \emph{RXTE} orbit fits where $K_{\mathrm{bb}}$
increases by as much as a factor $6.7$ (Figure~\ref{fig:best fit}).
$K_{\mathrm{bb}}$ is proportional to the black-body emitting area,
and the increase implies radius expansion, which is not expected to
take place in the decay phase of the superburst light curve. Also,
the count rate does not exhibit a dip, which is characteristic for
PRE \citep{Grindlay1980pre,Tawara1984,Lewin1984}. Furthermore, PRE
is typically observed for bursts from 4U\,1636--536 with peak black-body
fluxes in excess of $5\times10^{-8}\,\mathrm{erg\, s^{-1}\, cm^{-2}}$
\citep[in the PCA band;][]{Galloway2008catalog}, whereas the superburst
does not reach that flux (Figure~\ref{fig:flux}).

Similar increases in $K_{\mathrm{bb}}$ have been measured in the
tails of a number of PRE bursts from 4U\,1636--536, and are suggested
to originate from the deviation of the neutron star's photospheric
spectrum from a black body \citep[both studies assume the persistent flux is constant]{Guver2012,Zhang2013}.
\citet{Suleimanov2010} predict correction factors to $K_{\mathrm{bb}}$
that span a range of approximately a factor $8$ as a function of
flux. As the superburst peak does not reach the Eddington limit, however,
only a smaller range of a factor $\sim1.5$ is applicable. This can
only explain a small part of the increase in $K_{\mathrm{bb}}$. Alternatively,
variations in $K_{\mathrm{bb}}$ between bursts have been attributed
to different covering factors of the neutron star by the disk, depending
on its ionization state or on the geometry of the disk and boundary
layer \citep[e.g.,][]{Suleimanov2011}. So-called anisotropy factors
of up to $2$ have been predicted \citep{fujimoto88apj}, which falls
short of the increase that we observe.

We measure the black-body normalization in the first \emph{RXTE} orbit
as $K_{\mathrm{bb}}=62.3\pm0.4\,\mathrm{c\, s^{-1}\, cm^{-2}\, keV^{-1}}$.
At a distance of $6\,\mathrm{kpc}$ \citep[e.g.,][]{Galloway2008catalog},
this corresponds to a black-body radius of $4.736\pm0.009\,\mathrm{km}$
(not taking into account the uncertainty in the distance). The values
of $K_{\mathrm{bb}}$ measured for short non-PRE bursts from 4U~1636--536
that are observed with the PCA span a broad range, and the superburst
value is at the lower end of the distribution \citep[assuming a constant persistent flux]{Galloway2008catalog}.
In our analysis with a fixed persistent flux, $K_{\mathrm{bb}}$ was
$25\%$ larger compared to fits with a free normalization of the persistent
flux. The increased values of the normalization that we obtain in
fits where we keep $E_{\mathrm{cutoff}}$ fixed, span the range of
$K_{\mathrm{bb}}$ of the short bursts. As shown by \citet{Worpel2013}
for the case of PRE bursts, short X-ray bursts as well have an enhanced
persistent flux. The assumption of a constant flux leads, therefore,
to the over-prediction of the black body flux, which may be expressed
by an increased normalization.

We, therefore, regard it as unlikely that the increased black-body
emitting area is physical. The limited data quality does not allow
us to simultaneously constrain more parameters of the spectral components.
We choose to fix $K_{\mathrm{bb}}$ for the final three \emph{RXTE}
orbits, and allow the shape of the cut-off power law to change through
$E_{\mathrm{cutoff}}$. The $\chi^{2}$ values, however, suggest that
the fits are not optimal and the changes to the shape of the spectrum
require additional parameters. Nevertheless, the trends in the parameter
values appear to evolve smoothly between the first \emph{RXTE} orbit
and the subsequent orbits.

$E_{\mathrm{cutoff}}$ is substantially reduced with respect to the
pre-superburst value, and reaches a minimum around $t=7\times10^{3}\,\mathrm{s}$,
coinciding with a maximum in $N_{\mathrm{H}}$. This may correspond
to a viscous timescale on which the accretion disk responds to the
effects of the superburst. After this time, $E_{\mathrm{cutoff}}$
increases towards the pre-superburst value. A similar change in the
high energy tail of the spectrum has been noted for short bursts,
where the decrease of the count rate above $30\,\mathrm{keV}$ has
been linked to the cooling of the corona in response to the burst
\citep[e.g.,][]{Ji2014}.

\subsection{Strong local absorber}

We find a strong enhancement of the photoelectric absorption to improve
the fits: $N_{\mathrm{H}}$ is larger by an order of magnitude during
the superburst compared to the persistent emission. The absorption
component is typically used to describe interstellar absorption, but
this enhancement indicates the majority of the absorption during the
superburst is local to the binary. A local ionized absorber may provide
a more physically motivated spectral model. Furthermore, $N_{\mathrm{H}}$
is only constrained by a few spectral bins at low energy. One must,
therefore, use caution in interpreting our fit values for this parameter. 

A similar increase in $N_{\mathrm{H}}$ was observed for the 1999
superburst from 4U\,1820--30 \citep{Ballantyne2004}. For that superburst
it was speculated that the absorption was produced by material ejected
from the neutron star atmosphere during the PRE phase. 4U\,1636--536's
superburst, however, does not exhibit PRE. Alternatively, the absorbing
material may originate from a disk wind driven by the superburst emission
from the neutron star.

\subsection{Flux enhancement prior to the superburst}

At the start of the first \emph{RXTE }orbit of the superburst, a plateau
is visible where the flux is larger by a factor $1.46\pm0.03$ than
in the previous orbit. The spectrum of the plateau is not well-fit
by the persistent model (Section~\ref{sub:Persistent-emission:-fit}).
A broad excess remains at lower energies, which is well described
by a black body with an effective area that is consistent with the
value measured near the superburst peak (Figure~\ref{fig:best fit}).
This suggests that the neutron star surface has heated up prior to
the superburst. This is puzzling, because the precursor burst at $t=80\,\mathrm{s}$
is thought to be instigated by a shock within a second after the thermonuclear
runaway \citep{Weinberg2007,Keek2011}. Perhaps it is related to an
event that escaped detection because of the prior data gap. In two
cases a short Type I burst has been observed within $30$~minutes
of the superburst onset, and at both times the flux remained at a
somewhat higher value until the superburst rise \citep{Kuulkers2002ks1731,Chenevez2011ATel}.
Because of the small number of occurrences it is, however, not excluded
that these short bursts merely happened by chance, and are unrelated
to the superburst ignition.

\subsection{Comparison of analyses and superbursts}

Our results from the classical fit with a fixed persistent flux are
consistent with the preliminary analysis reported by \citet{2004Kuulkers}
as well as the analysis of the ASM data of the superburst \citep{Wijnands2001sb}.
Our ``optimal'' spectral model with variable persistent flux yields
substantial contributions to the flux from both the black body and
the cut-off power law. We use the black-body flux to measure the emission
from the neutron star surface due to thermonuclear burning. For the
reasons explained in Section~\ref{sub:The-persistent-spectrum},
we do not expect a large contribution from a boundary layer to the
black body, especially near the superburst peak. The enhancement of
the cut-off power law suggests that it is powered by the star's surface
emission. This, however, must be black-body radiation that was not
in our direct line of sight, and that is accounted for if we use only
the black body and assume isotropic emission (see also \citealt{2002Kuulkers}
for a discussion on decoupling burst and persistent flux). 

\begin{table}
\caption{\label{tab:Superburst-parameters-from}Burst properties: classical
spectral fit vs optimal fit}

\begin{tabular}{cll}
\hline 
Spectral model: & Classical\,$^{\mathrm{a}}$ & Optimal\,$^{\mathrm{b}}$\tabularnewline
\hline 
$t_{\mathrm{peak}}(\mathrm{s})$$^{\mathrm{c}}$ & $928$ & $864$\tabularnewline
$F_{\mathrm{peak}}\,(\mathrm{10^{-8}erg\, s^{-1}})$  & $2.421\pm0.005$ & $2.19\pm0.06$\tabularnewline
$kT_{\mathrm{peak}}\,(\mathrm{keV})$ & $2.278\pm0.004$ & $2.386\pm0.012$\tabularnewline
$K_{\mathrm{bb,peak}}\,(\mathrm{c\, s^{-1}\, cm^{-2}\, keV^{-1}})$ & $92.0\pm0.6$ & $69\pm3$\tabularnewline
Fluence~$(10^{-4}\mathrm{erg})$ & $1.458\pm0.006$ & $1.10\pm0.09$\tabularnewline
\hline 
\end{tabular}

$^{\mathrm{a}}$ Section~\ref{sub:Classical-superburst-spectral}.
Note: uncertainties in parameters do not include a correction for
the high $\chi^{2}$ in this fit.

$^{\mathrm{b}}$ Section~\ref{sub:orbit 2}

$^{\mathrm{c}}$ The parameters are for the black body in the $3$
to $20\,\mathrm{keV}$ range at the time of maximal flux, $t_{\mathrm{peak}}$:
unabsorbed flux , $F_{\mathrm{peak}}$; temperature, $kT_{\mathrm{peak}}$;
normalization, $K_{\mathrm{bb,peak}}$. The $3$ to $20\,\mathrm{keV}$
unabsorbed black-body fluence includes linear interpolation over the
data gaps.\\
~
\end{table}
Using the optimal spectral model gives different results for some
superburst properties compared to the classical fit (Table~\ref{tab:Superburst-parameters-from}).
The peak times are consistent: $t_{\mathrm{peak}}$ is different by
only one time bin of the spectra. The peak black-body temperature
differs by $10\%$. The peak flux is lower in our optimal model, because
a smaller part of the flux is attributed to the black-body component.
This, combined with the higher temperature, leads to a lower black-body
normalization. The largest difference is, therefore, apparent in the
measured black-body normalization.

Comparison of the classical fit to other superbursts, characterizes
this superburst as ``typical'' \citep[e.g.,][]{Kuulkers2003a,Keek2008int..work}.
The increase of the persistent flux and its effect on the derived
burst properties, may play a role in other superbursts as well, although
usually the data is of insufficient quality to determine this. If
the superburst fluence is smaller than previously derived, the inferred
energy content of the fuel is lower. Using our values of the fluence
(Table~\ref{tab:Superburst-parameters-from}), this means that the
amount of carbon or of heavy \textsl{rp}-process ashes \citep[targets for photodisintegration;][]{Schatz2003ApJ}
in the neutron star ocean is lower by $25\%\pm2\%$ \citep{Cumming2001,Cumming2006}.
As it has been challenging to explain the production of enough carbon
to power superbursts \citep[e.g.,][]{Woosley2004}, this reduction
alleviates the problem, although it does not completely solve it.
In a forthcoming paper we will investigate the consequences using
numerical models of the nuclear burning in the neutron star envelope.

\section{Conclusions}

Detailed time resolved spectral analysis of the 2001 superburst from
4U\,1636-536 reveals an emission line and absorption edge suggestive
of reflection of the superburst off the accretion disk. This is only
the second superburst for which this has been observed, and shows
that typical non-PRE superbursts also are powerful enough to produce
these features. Furthermore, we find an increase in the persistent
flux during the burst. We observe the shape of the non-thermal component
of the spectrum to change under influence of the superburst, and return
to the pre-superburst state in the burst tail. The increase in the
persistent flux may suggest that previous measurements overestimate
the energetics of superbursts. This has consequences for the inference
of the composition of the neutron star ocean that is the fuel for
superbursts. In forthcoming papers we will discuss how this changes
our understanding of nuclear burning in the envelope, and we will
employ detailed models to study the reflection spectrum.

\acknowledgements{LK and DRB acknowledge support from NASA ADAP grant NNX13AI47G and
NSF award AST 1008067.}

\bibliographystyle{apj}
\bibliography{apj-jour,sbspectr}

\begin{thebibliography}{}
\expandafter\ifx\csname natexlab\endcsname\relax\def\natexlab#1{#1}\fi

\bibitem[{{Altamirano} {et~al.}(2008){Altamirano}, {van der Klis}, {Wijnands},
  \& {Cumming}}]{Altamirano2008}
{Altamirano}, D., {van der Klis}, M., {Wijnands}, R., \& {Cumming}, A. 2008,
  \apjl, 673, L35

\bibitem[{{Arnaud}(1996)}]{Arnaud1996}
{Arnaud}, K.~A. 1996, in ASP Conf. Ser. 101: Astronomical Data Analysis
  Software and Systems V, ed. G.~H. {Jacoby} \& J.~{Barnes}, 17

\bibitem[{{Ballantyne} \& {Everett}(2005)}]{Ballantyne2005}
{Ballantyne}, D.~R., \& {Everett}, J.~E. 2005, \apj, 626, 364

\bibitem[{{Ballantyne} \& {Strohmayer}(2004)}]{Ballantyne2004}
{Ballantyne}, D.~R., \& {Strohmayer}, T.~E. 2004, \apjl, 602, L105

\bibitem[{{Balucinska-Church} \& {McCammon}(1992)}]{Balucinska1992}
{Balucinska-Church}, M., \& {McCammon}, D. 1992, \apj, 400, 699

\bibitem[{{Bhattacharyya} \& {Strohmayer}(2006)}]{Bhattacharyya2006}
{Bhattacharyya}, S., \& {Strohmayer}, T.~E. 2006, \apjl, 636, L121

\bibitem[{{Bhattacharyya} \& {Strohmayer}(2007)}]{Bhattacharyya2007}
---. 2007, \apjl, 664, L103

\bibitem[{{Bradt} {et~al.}(1993){Bradt}, {Rothschild}, \& {Swank}}]{Bradt1993}
{Bradt}, H.~V., {Rothschild}, R.~E., \& {Swank}, J.~H. 1993, \aaps, 97, 355

\bibitem[{{Cackett} {et~al.}(2008){Cackett}, {Miller}, {Bhattacharyya},
  {Grindlay}, {Homan}, {van der Klis}, {Miller}, {Strohmayer}, \&
  {Wijnands}}]{Cackett2008}
{Cackett}, E.~M., {Miller}, J.~M., {Bhattacharyya}, S., {et~al.} 2008, \apj,
  674, 415

\bibitem[{{Chen} {et~al.}(2013){Chen}, {Zhang}, {Zhang}, {Ji}, {Torres},
  {Kretschmar}, {Li}, \& {Wang}}]{Chen2013}
{Chen}, Y.-P., {Zhang}, S., {Zhang}, S.-N., {et~al.} 2013, \apjl, 777, L9

\bibitem[{{Chen} {et~al.}(2012){Chen}, {Zhang}, {Zhang}, {Li}, \&
  {Wang}}]{Chen2012}
{Chen}, Y.-P., {Zhang}, S., {Zhang}, S.-N., {Li}, J., \& {Wang}, J.-M. 2012,
  \apjl, 752, L34

\bibitem[{{Chenevez} {et~al.}(2011){Chenevez}, {Brandt}, {Kuulkers},
  {Alfonso-Garzon}, {Beckmann}, {Bird}, {Courvoisier}, {Del Santo}, {Domingo},
  {Ebisawa}, {Jonker}, {Kretschmar}, {Markwardt}, {Oosterbroek}, {Paizis},
  {Pottschmidt}, {Sanchez-Fernandez}, \& {Wijnands}}]{Chenevez2011ATel}
{Chenevez}, J., {Brandt}, S., {Kuulkers}, E., {et~al.} 2011, The Astronomer's
  Telegram, 3183, 1

\bibitem[{{Cooper} {et~al.}(2009){Cooper}, {Steiner}, \& {Brown}}]{Cooper2009}
{Cooper}, R.~L., {Steiner}, A.~W., \& {Brown}, E.~F. 2009, \apj, 702, 660

\bibitem[{{Cornelisse} {et~al.}(2000){Cornelisse}, {Heise}, {Kuulkers},
  {Verbunt}, \& {in~'t~Zand}}]{Cornelisse2000}
{Cornelisse}, R., {Heise}, J., {Kuulkers}, E., {Verbunt}, F., \& {in~'t~Zand},
  J.~J.~M. 2000, \aap, 357, L21

\bibitem[{{Cumming} \& {Bildsten}(2001)}]{Cumming2001}
{Cumming}, A., \& {Bildsten}, L. 2001, \apjl, 559, L127

\bibitem[{{Cumming} {et~al.}(2006){Cumming}, {Macbeth}, {in~'t~Zand}, \&
  {Page}}]{Cumming2006}
{Cumming}, A., {Macbeth}, J., {in~'t~Zand}, J.~J.~M., \& {Page}, D. 2006, \apj,
  646, 429

\bibitem[{{Fujimoto}(1988)}]{fujimoto88apj}
{Fujimoto}, M.~Y. 1988, \apj, 324, 995

\bibitem[{{Galloway} {et~al.}(2008){Galloway}, {Muno}, {Hartman}, {Psaltis}, \&
  {Chakrabarty}}]{Galloway2008catalog}
{Galloway}, D.~K., {Muno}, M.~P., {Hartman}, J.~M., {Psaltis}, D., \&
  {Chakrabarty}, D. 2008, \apjs, 179, 360

\bibitem[{{Grindlay} {et~al.}(1980){Grindlay}, {Marshall}, {Hertz},
  {Weisskopf}, {Elsner}, {Ghosh}, {Darbro}, {Sutherland}, \&
  {Soltan}}]{Grindlay1980pre}
{Grindlay}, J.~E., {Marshall}, H.~L., {Hertz}, P., {et~al.} 1980, \apjl, 240,
  L121

\bibitem[{{G{\"u}ver} {et~al.}(2012){G{\"u}ver}, {Psaltis}, \&
  {{\"O}zel}}]{Guver2012}
{G{\"u}ver}, T., {Psaltis}, D., \& {{\"O}zel}, F. 2012, \apj, 747, 76

\bibitem[{{Hoffman} {et~al.}(1977){Hoffman}, {Lewin}, \& {Doty}}]{Hoffman1977}
{Hoffman}, J.~A., {Lewin}, W.~H.~G., \& {Doty}, J. 1977, \apjl, 217, L23

\bibitem[{{in~'t~Zand} {et~al.}(2011){in~'t~Zand}, {Galloway}, \&
  {Ballantyne}}]{Zand2011}
{in~'t~Zand}, J.~J.~M., {Galloway}, D.~K., \& {Ballantyne}, D.~R. 2011, \aap,
  525, A111

\bibitem[{{in~'t Zand} \& {Weinberg}(2010)}]{Zand2010}
{in~'t Zand}, J.~J.~M., \& {Weinberg}, N.~N. 2010, \aap, 520, A81

\bibitem[{{in~'t Zand} {et~al.}(2013){in~'t Zand}, {Galloway}, {Marshall},
  {Ballantyne}, {Jonker}, {Paerels}, {Palmer}, {Patruno}, \&
  {Weinberg}}]{Zand2013}
{in~'t Zand}, J.~J.~M., {Galloway}, D.~K., {Marshall}, H.~L., {et~al.} 2013,
  \aap, 553, A83

\bibitem[{{Jahoda} {et~al.}(2006){Jahoda}, {Markwardt}, {Radeva}, {Rots},
  {Stark}, {Swank}, {Strohmayer}, \& {Zhang}}]{Jahoda2006}
{Jahoda}, K., {Markwardt}, C.~B., {Radeva}, Y., {et~al.} 2006, \apjs, 163, 401

\bibitem[{{Ji} {et~al.}(2014){Ji}, {Zhang}, {Chen}, {Zhang}, {Torres},
  {Kretschmar}, \& {Li}}]{Ji2014}
{Ji}, L., {Zhang}, S., {Chen}, Y., {et~al.} 2014, \apj, 782, 40

\bibitem[{{Keek}(2012)}]{Keek2012precursors}
{Keek}, L. 2012, \apj, 756, 130

\bibitem[{{Keek} {et~al.}(2010){Keek}, {Galloway}, {in 't Zand}, \&
  {Heger}}]{Keek2010}
{Keek}, L., {Galloway}, D.~K., {in 't Zand}, J.~J.~M., \& {Heger}, A. 2010,
  \apj, 718, 292

\bibitem[{{Keek} \& {Heger}(2011)}]{Keek2011}
{Keek}, L., \& {Heger}, A. 2011, \apj, 743, 189

\bibitem[{{Keek} {et~al.}(2012){Keek}, {Heger}, \& {in't Zand}}]{Keek2012}
{Keek}, L., {Heger}, A., \& {in't Zand}, J.~J.~M. 2012, \apj, 752, 150

\bibitem[{{Keek} \& {in~'t~Zand}(2008)}]{Keek2008int..work}
{Keek}, L., \& {in~'t~Zand}, J.~J.~M. 2008, in Proceedings of the 7th INTEGRAL
  Workshop. 8 - 11 September 2008 Copenhagen, Denmark. Online at
  http://pos.sissa.it/cgi-bin/reader/conf.cgi?confid=67, p.32

\bibitem[{Kuulkers(2004)}]{Kuulkers2003a}
Kuulkers, E. 2004, Nucl. Phys. Proc. Suppl., 132, 466

\bibitem[{{Kuulkers}(2009)}]{Kuulkers2009ATel}
{Kuulkers}, E. 2009, The Astronomer's Telegram, 2140, 1

\bibitem[{{Kuulkers} {et~al.}(2002{\natexlab{a}}){Kuulkers}, {Homan}, {van der
  Klis}, {Lewin}, \& {M{\'e}ndez}}]{2002Kuulkers}
{Kuulkers}, E., {Homan}, J., {van der Klis}, M., {Lewin}, W.~H.~G., \&
  {M{\'e}ndez}, M. 2002{\natexlab{a}}, \aap, 382, 947

\bibitem[{{Kuulkers} {et~al.}(2004){Kuulkers}, {in~'t~Zand}, {Homan}, {van
  Straaten}, {Altamirano}, \& {van der Klis}}]{2004Kuulkers}
{Kuulkers}, E., {in~'t~Zand}, J., {Homan}, J., {et~al.} 2004, in AIP Conf.
  Proc. 714: X-ray Timing 2003: Rossi and Beyond, 257--260

\bibitem[{{Kuulkers} {et~al.}(2002{\natexlab{b}}){Kuulkers}, {in~'t~Zand}, {van
  Kerkwijk}, {Cornelisse}, {Smith}, {Heise}, {Bazzano}, {Cocchi}, {Natalucci},
  \& {Ubertini}}]{Kuulkers2002ks1731}
{Kuulkers}, E., {in~'t~Zand}, J.~J.~M., {van Kerkwijk}, M.~H., {et~al.}
  2002{\natexlab{b}}, \aap, 382, 503

\bibitem[{{Kuulkers} {et~al.}(2010){Kuulkers}, {in 't Zand}, {Atteia},
  {Levine}, {Brandt}, {Smith}, {Linares}, {Falanga},
  {S{\'a}nchez-Fern{\'a}ndez}, {Markwardt}, {Strohmayer}, {Cumming}, \&
  {Suzuki}}]{Kuulkers2010}
{Kuulkers}, E., {in 't Zand}, J.~J.~M., {Atteia}, J., {et~al.} 2010, \aap, 514,
  A65+

\bibitem[{{Levine} {et~al.}(1996){Levine}, {Bradt}, {Cui}, {Jernigan},
  {Morgan}, {Remillard}, {Shirey}, \& {Smith}}]{Levine1996}
{Levine}, A.~M., {Bradt}, H., {Cui}, W., {et~al.} 1996, \apjl, 469, L33+

\bibitem[{{Lewin} {et~al.}(1984){Lewin}, {Vacca}, \& {Basinska}}]{Lewin1984}
{Lewin}, W.~H.~G., {Vacca}, W.~D., \& {Basinska}, E.~M. 1984, \apjl, 277, L57

\bibitem[{{Lyu} {et~al.}(2014){Lyu}, {Mendez}, {Sanna}, {Homan}, {Belloni}, \&
  {Hiemstra}}]{Lyu2014}
{Lyu}, M., {Mendez}, M., {Sanna}, A., {et~al.} 2014, ArXiv e-prints,
  arXiv:1402.2350

\bibitem[{{Maccarone} \& {Coppi}(2003)}]{Maccarone2003}
{Maccarone}, T.~J., \& {Coppi}, P.~S. 2003, \aap, 399, 1151

\bibitem[{{Miller}(2007)}]{Miller2007}
{Miller}, J.~M. 2007, \araa, 45, 441

\bibitem[{{Miller} {et~al.}(2013){Miller}, {Parker}, {Fuerst}, {Bachetti},
  {Barret}, {Grefenstette}, {Tendulkar}, {Harrison}, {Boggs}, {Chakrabarty},
  {Christensen}, {Craig}, {Fabian}, {Hailey}, {Natalucci}, {Paerels}, {Rana},
  {Stern}, {Tomsick}, \& {Zhang}}]{Miller2013}
{Miller}, J.~M., {Parker}, M.~L., {Fuerst}, F., {et~al.} 2013, \apjl, 779, L2

\bibitem[{{Negoro} {et~al.}(2012){Negoro}, {Asada}, {Serino}, {Nakahira},
  {Morii}, {Ogawa}, {Ueno}, {Tomida}, {Ishikawa}, {Yamaoka}, {Kimura},
  {Mihara}, {Sugizaki}, {Morihana}, {Yamamoto}, {Sugimoto}, {Takagi},
  {Matsuoka}, {Kawai}, {Usui}, {Ishikawa}, {Yoshida}, {Sakamoto}, {Nakano},
  {Tsunemi}, {Sasaki}, {Nakajima}, {Ueda}, {Hiroi}, {Shidatsu}, {Sato},
  {Kawamuro}, {Tsuboi}, {Yamauchi}, {Nishimura}, {Hanayama}, \&
  {Yoshidom}}]{Negoro2012}
{Negoro}, H., {Asada}, M., {Serino}, M., {et~al.} 2012, The Astronomer's
  Telegram, 4622, 1

\bibitem[{{Pandel} {et~al.}(2008){Pandel}, {Kaaret}, \& {Corbel}}]{Pandel2008}
{Pandel}, D., {Kaaret}, P., \& {Corbel}, S. 2008, \apj, 688, 1288

\bibitem[{{Revnivtsev} {et~al.}(2001){Revnivtsev}, {Churazov}, {Gilfanov}, \&
  {Sunyaev}}]{Revnivtsev2001}
{Revnivtsev}, M., {Churazov}, E., {Gilfanov}, M., \& {Sunyaev}, R. 2001, \aap,
  372, 138

\bibitem[{{Rothschild} {et~al.}(1998){Rothschild}, {Blanco}, {Gruber},
  {Heindl}, {MacDonald}, {Marsden}, {Pelling}, {Wayne}, \&
  {Hink}}]{Rothschild1998}
{Rothschild}, R.~E., {Blanco}, P.~R., {Gruber}, D.~E., {et~al.} 1998, \apj,
  496, 538

\bibitem[{{Schatz} {et~al.}(2003){Schatz}, {Bildsten}, \&
  {Cumming}}]{Schatz2003ApJ}
{Schatz}, H., {Bildsten}, L., \& {Cumming}, A. 2003, \apjl, 583, L87

\bibitem[{{Strohmayer} \& {Brown}(2002)}]{Strohmayer2002}
{Strohmayer}, T.~E., \& {Brown}, E.~F. 2002, \apj, 566, 1045

\bibitem[{{Strohmayer} \& {Markwardt}(2002)}]{Strohmayer2002a}
{Strohmayer}, T.~E., \& {Markwardt}, C.~B. 2002, \apj, 577, 337

\bibitem[{{Strohmayer} {et~al.}(1998){Strohmayer}, {Zhang}, {Swank}, {White},
  \& {Lapidus}}]{Strohmayer1998}
{Strohmayer}, T.~E., {Zhang}, W., {Swank}, J.~H., {White}, N.~E., \& {Lapidus},
  I. 1998, \apjl, 498, L135+

\bibitem[{{Suleimanov} {et~al.}(2011{\natexlab{a}}){Suleimanov}, {Poutanen},
  {Revnivtsev}, \& {Werner}}]{Suleimanov2011}
{Suleimanov}, V., {Poutanen}, J., {Revnivtsev}, M., \& {Werner}, K.
  2011{\natexlab{a}}, \apj, 742, 122

\bibitem[{{Suleimanov} {et~al.}(2011{\natexlab{b}}){Suleimanov}, {Poutanen}, \&
  {Werner}}]{Suleimanov2010}
{Suleimanov}, V., {Poutanen}, J., \& {Werner}, K. 2011{\natexlab{b}}, \aap,
  527, A139+

\bibitem[{{Swank} {et~al.}(1977){Swank}, {Becker}, {Boldt}, {Holt}, {Pravdo},
  \& {Serlemitsos}}]{swank1977}
{Swank}, J.~H., {Becker}, R.~H., {Boldt}, E.~A., {et~al.} 1977, \apjl, 212, L73

\bibitem[{{Tawara} {et~al.}(1984){Tawara}, {Kii}, {Hayakawa}, {Kunieda},
  {Masai}, {Nagase}, {Inoue}, {Koyama}, {Makino}, {Makishima}, {Matsuoka},
  {Murakami}, {Oda}, {Ogawara}, {Ohashi}, {Shibazaki}, {Tanaka}, {Miyamoto},
  {Tsunemi}, {Yamashita}, \& {Kondo}}]{Tawara1984}
{Tawara}, Y., {Kii}, T., {Hayakawa}, S., {et~al.} 1984, \apjl, 276, L41

\bibitem[{{van Paradijs} \& {Lewin}(1986)}]{Paradijs1986spectra}
{van Paradijs}, J., \& {Lewin}, H.~G. 1986, \aap, 157, L10

\bibitem[{{van Paradijs} {et~al.}(1986){van Paradijs}, {Sztajno}, {Lewin},
  {Trumper}, {Vacca}, \& {van der Klis}}]{Paradijs1986}
{van Paradijs}, J., {Sztajno}, M., {Lewin}, W.~H.~G., {et~al.} 1986, \mnras,
  221, 617

\bibitem[{{Walker}(1992)}]{Walker1992}
{Walker}, M.~A. 1992, \apj, 385, 642

\bibitem[{{Weinberg} \& {Bildsten}(2007)}]{Weinberg2007}
{Weinberg}, N.~N., \& {Bildsten}, L. 2007, \apj, 670, 1291

\bibitem[{{Wijnands}(2001)}]{Wijnands2001sb}
{Wijnands}, R. 2001, \apjl, 554, L59

\bibitem[{{Wilms} {et~al.}(2000){Wilms}, {Allen}, \& {McCray}}]{Wilms2000}
{Wilms}, J., {Allen}, A., \& {McCray}, R. 2000, \apj, 542, 914

\bibitem[{{Woosley} {et~al.}(2004){Woosley}, {Heger}, {Cumming}, {Hoffman},
  {Pruet}, {Rauscher}, {Fisker}, {Schatz}, {Brown}, \&
  {Wiescher}}]{Woosley2004}
{Woosley}, S.~E., {Heger}, A., {Cumming}, A., {et~al.} 2004, \apjs, 151, 75

\bibitem[{{Worpel} {et~al.}(2013){Worpel}, {Galloway}, \& {Price}}]{Worpel2013}
{Worpel}, H., {Galloway}, D.~K., \& {Price}, D.~J. 2013, ArXiv e-prints,
  arXiv:1303.4824

\bibitem[{{Zhang} {et~al.}(2009){Zhang}, {M{\'e}ndez}, {Altamirano}, {Belloni},
  \& {Homan}}]{Zhang2009}
{Zhang}, G., {M{\'e}ndez}, M., {Altamirano}, D., {Belloni}, T.~M., \& {Homan},
  J. 2009, \mnras, 398, 368

\bibitem[{{Zhang} {et~al.}(2013){Zhang}, {M{\'e}ndez}, {Belloni}, \&
  {Homan}}]{Zhang2013}
{Zhang}, G., {M{\'e}ndez}, M., {Belloni}, T.~M., \& {Homan}, J. 2013, \mnras,
  436, 2276

\end{thebibliography}

\end{document}